\documentclass[10pt,conference,twocolumn]{IEEEtran}

\usepackage{amsmath}
\usepackage{amssymb}
\usepackage{latexsym}
\usepackage{amsfonts}

\begin{document}
\pagestyle{headings}  % switches on printing of running heads
\vspace{1cm}

\title{On discrete cosine transform}

\author{
\authorblockN{Jianqin Zhou}
\authorblockA{ 1.Telecommunication School, Hangzhou Dianzi University,
Hangzhou, 310018 China\\ 2.Computer Science School, Anhui Univ. of
Technology, Ma'anshan, 243002 China\\ \ \ zhou9@yahoo.com
 }
}
\maketitle              % typeset the title of the contribution

\begin{abstract}
The discrete cosine transform (DCT), introduced by Ahmed, Natarajan
and Rao, has been used in many applications of digital signal
processing, data compression and information hiding. There are four
types of the discrete cosine transform. In simulating the discrete
cosine transform, we propose a generalized discrete cosine transform
with three parameters, and prove its orthogonality for some new
cases. A new type of discrete cosine transform is proposed and its
orthogonality is proved. Finally,  we propose a generalized discrete
W transform with three parameters, and prove its orthogonality for
some new cases.

\end{abstract}

\section{Introduction}
Discrete Fourier transform has been an important tool in many
applications of digital signal processing, image processing and
information hiding. The appearance of fast fourier transform (FFT)
has greatly promoted the rapid development of the subjects above.  Ahmed, Natarajan, and Rao (1974) %\cite{Ahmed} 
proposed discrete
cosine transform defined on real number field, it can be called
DCT-II-E or DCT-III-E (Wang and Hunt,1983). 
Jain (1974)%\cite{Jain}
proposed discrete fourier transform DCT-IV-E, and  Wang and
Hunt (1983)%\cite{Wang, Wang84}
 proposed discrete cosine transform DCT-I-E.
The discrete cosine transform has been used in frequency spectrum
analysis, data compression, convolution computation and information
hiding. Its theory and algorithms have  received much attention for
the last two decades (Wang and Hunt, 1983, Wang, 1984, August,2004, Kunz,2008).

It is demonstrated that the performance of discrete cosine transform
can well approximate to ideal K-L transform (Karhunen-Loeve
Transform) (Ahmed,Natarajan and Rao,1974). 
K-L transform was proposed to dealing with a
class of extensive stochastic image. After the image being
transformed with K-L transform, the image restored from the result
is the best approximation to the original image in the statistical
sense. Moreover, for the common data model of Markov process, when
the correlation coefficient $r=1$, K-L transform is degraded to the
classic DCT transform. In fact, Real-world images are neither
stationary nor Markovian. They have  different textures and
structures, important image structures like edges, arris and lines
extend over large distances in the image (Kunz,2008). Therefore,
different types of transformation are desirable to meet the
different applications.

Actually, the discrete cosine transform (DCT-III-E) can be
generalized to the  unified form with  parameters $p$, $q$ and $r$,
as follows:
\begin{eqnarray*}
X(k)=\alpha(k)\sum\limits^{N-1}_{n=0}x(n)\cos\frac{k(4qn+r)p\pi}{2N},\\
k=0,1,\cdots,N-1.
\end{eqnarray*}

There are many new meaningful transforms, such as when
$(p,q,r)=(1,1,1)$, the new transform is  orthogonal. Generally, when
$\gcd(pq,N)=1$, $\gcd(pr,2)=1$, where $p$, $q$ and $r$ are  positive
integers, the new transform is orthogonal.

In a similar way,  we generalize discrete cosine transforms DCT-II-E
and DCT-IV-E. Furthermore, a new type of  discrete cosine transform,
a new type of  discrete sine transform and a new type of discrete
sine-cosine transform are proposed, and their orthogonality are
proved.

The discrete W transform (DWT), introduced by Wang Z, has been used
in many applications of digital signal processing, data compression
and information hiding. The unified form of discrete W transform has
two  parameters. There are four useful types of the discrete W
transform. In simulating the discrete W transform, we propose a
generalized discrete W transform with three parameters, and prove
its orthogonality for some new cases.

\section{Generalized discrete cosine transform}

Let $\{x(n);n=0,1,2,\cdots$, $N\}$ be a vector of real numbers. The
definitions of four common types of  discrete cosine transform
(Wang and Hunt,1983) are given as follows:

DCT-I-E:
\begin{eqnarray*}
X(k)=\sqrt{\frac{2}{N}}\alpha(k)\sum\limits^{N}_{n=0}\alpha(n)x(n)\cos\frac{kn\pi}{N},\\
k=0,1,\cdots,N,
\end{eqnarray*}
where
\begin{eqnarray*}&&\alpha(k)=\left\{
\begin{array}{ll}
\frac{1}{\sqrt 2} &             k=0\mbox{ or }N\\
\ \ 1               &          \mbox{else}\\
\end{array}\right.\\
&&\alpha(n)=\left\{
\begin{array}{ll}
\frac{1}{\sqrt 2} &             n=0\mbox{ or }N\\
\ \ 1               &            \mbox{else}\\
\end{array}\right.
\end{eqnarray*}

DCT-II-E:
\begin{eqnarray*}X(k)=\sum\limits^{N-1}_{n=0}\alpha(n)x(n)\cos\frac{(2k+1)n\pi}{2N},\\
k=0,1,\cdots,N-1,
\end{eqnarray*}
 where $\alpha(n)=\left\{
\begin{array}{ll}
\sqrt\frac{1}{N} &             n=0\\
\sqrt\frac{2}{N} &            \mbox{else}\\
\end{array}\right.$

DCT-III-E:
\begin{eqnarray*}
X(k)=\alpha(k)\sum\limits^{N-1}_{n=0}x(n)\cos\frac{k(2n+1)\pi}{2N},\\
k=0,1,\cdots,N-1,
\end{eqnarray*}
 where $ \alpha(k)=\left\{
\begin{array}{ll}
\sqrt\frac{1}{N} &             k=0\\
\sqrt\frac{2}{N} &            \mbox{else}\\
\end{array}\right.$

DCT-IV-E:
\begin{eqnarray*}
X(k)=\sqrt{\frac{2}{N}}\sum\limits^{N-1}_{n=0}x(n)\cos\frac{(2k+1)(2n+1)\pi}{2N},\\
k=0,1,\cdots,N-1.
\end{eqnarray*}

Actually, the discrete cosine transform (DCT-III-E) can be
generalized to the following unified form with  parameters $p$, $q$
and $r$:
\begin{eqnarray}
X(k) =\alpha(k)\sum\limits^{N-1}_{n=0}x(n)\cos\frac{k(4qn+r)p\pi}{2N},\label{formula01}\\
k=0,1,\cdots,N-1.\notag
\end{eqnarray}

We now prove that when $\gcd(pq,N)=1$, $\gcd(pr,2)=1$, where $p$,
$q$ and $r$ are positive integers,  transform (\ref{formula01}) is
orthogonal.

Transform (\ref{formula01}) can be written into matrix form:
\begin{eqnarray*}
X(N)=C(N)\cdot x(N)
\end{eqnarray*}
where $X(N), x(N)$ are column vectors of length $N,$

%\onecolumn

 {\tiny
$ C(N)=\sqrt{\frac{2}{N}}\\ \left(
\begin{array}{cccccllllll}
\frac{1}{\sqrt{2}}        &\frac{1}{\sqrt{2}}      &\frac{1}{\sqrt{2}}     &\cdots  & \frac{1}{\sqrt{2}}\\
\cos\frac{rp\pi}{2N}   &\cos\frac{(4q+r)p\pi}{2N}   &\cos\frac{(8q+r)p\pi}{2N} &\cdots  & \cos\frac{[4q(N-1)+r]p\pi}{2N}\\
\cos\frac{2rp\pi}{2N}  &\cos\frac{2(4q+r)p\pi}{2N}   &\cos\frac{2(8q+r)p\pi}{2N} &\cdots  & \cos\frac{2[4q(N-1)+r]p\pi}{2N}\\
\vdots&                \vdots                 &\vdots&     \vdots&       \vdots \\
\cos\frac{\eta r}{2N}   &\cos\frac{\eta (4q+r)}{2N}   &\cos\frac{\eta (8q+r)}{2N} &\cdots  & \cos\frac{\eta [4q(N-1)+r]}{2N}\\
\end{array}\right)$
}\\\\
where $\eta =(N-1)p\pi$

%\twocolumn

 To prove that the new transform (1) is orthogonal,  or equivalently, to prove that $C(N)$ is an orthogonal
 matrix. For the sake of simplicity, the coefficients
 $\sqrt{\frac{2}{N}}$ and $\frac{1}{\sqrt {2}}$  is omitted.

Let $0\leq k_1<k_2<N$, the inner product of the $k_1$th row and
$k_2$th row of $C(N)$ is that£º
\begin{eqnarray}
&&\cos\frac{k_1rp\pi}{2N}\cos\frac{k_2rp\pi}{2N}\notag\\
&&+\cos\frac{k_1(4q+r)p\pi}{2N}\cos\frac{k_2(4q+r)p\pi}{2N}+\cdots\notag\\
&&+\cos\frac{k_1[4q(N-1)+r]p\pi}{2N}\cos\frac{k_2[4q(N-1)+r]p\pi}{2N}\notag\\
&=&\frac{1}{2}\{\cos\frac{(k_1-k_2)rp\pi}{2N}+\cos\frac{(k_1+k_2)rp\pi}{2N}\notag\\
&&+\cos\frac{(k_1-k_2)(4q+r)p\pi}{2N}+\cos\frac{(k_1+k_2)(4q+r)p\pi}{2N}\notag\\
&&+\cdots+\cos\frac{(k_1-k_2)[4q(N-1)+r]p\pi}{2N}\notag\\
&&+\cos\frac{(k_1+k_2)[4q(N-1)+r]p\pi}{2N}\}\label{formula02}
\end{eqnarray}
where the above equality follows by $$\cos A\cos B=\frac{1}{2}[\cos
(A-B)+\cos(A+B)]$$

Since  $e^{ix}=\cos x+i\sin x$, we consider the real part of the
following complex number:
\begin{eqnarray*}
&& e^{i\frac{krp\pi}{2N}}+e^{i\frac{k(4q+r)p\pi}{2N}}+\cdots+e^{i\frac{k[4q(N-1)+r]p\pi}{2N}}\\
&=&\frac{e^{i\frac{krp\pi}{2N}}}{1-e^{i\frac{2kqp\pi}{N}}}(1-e^{i\frac{2kqp\pi
N}{N}})
\end{eqnarray*}

Note that $\gcd(pq,N)=1$ and $\gcd(pr,2)=1$. When $0 < k < 2N$, if
$1-e^{i\frac{2kpq\pi}{N}}=0$, then $k=N$, the real part of

$$e^{i\frac{krp\pi}{2N}}+e^{i\frac{k(4q+r)p\pi}{2N}}+\cdots+e^{i\frac{k[4q(N-1)+r]p\pi}{2N}}$$
is zero.

If $1-e^{i\frac{2kpq\pi}{N}}\neq0$, then
$$e^{i\frac{krp\pi}{2N}}+e^{i\frac{k(4q+r)p\pi}{2N}}+\cdots+e^{i\frac{k[4q(N-1)+r]p\pi}{2N}}=0$$

From equality (\ref{formula02}), when $0\leq k_1<k_2<N$, the inner
product of the $k_1$th row and $k_2$th row of $C(N)$ is zero.  We
know that $C(N)$
is an orthogonal matrix.\\

Let $0< k_1<N$, the inner product of the $k_1$th row and itself of
$C(N)$ is that£º
\begin{eqnarray}
&&\frac{2}{N}\{(\cos\frac{k_1rp\pi}{2N})^2+[\cos\frac{k_1(4q+r)p\pi}{2N}]^2+\cdots\notag\\
&&+[\cos\frac{k_1[4q(N-1)+r]p\pi}{2N}]^2\}\notag\\
&=&1+\frac{1}{N}\{\cos\frac{k_1rp\pi}{N}+\cos\frac{k_1(4q+r)p\pi}{N}\notag\\
&&+\cdots+\cos\frac{k_1[4q(N-1)+r]p\pi}{N}\}\label{formula020}
\end{eqnarray}

Note that
\begin{eqnarray*}
&& e^{i\frac{krp\pi}{N}}+e^{i\frac{k(4q+r)p\pi}{N}}+\cdots+e^{i\frac{k[4q(N-1)+r]p\pi}{N}}\\
&=&\frac{e^{i\frac{krp\pi}{N}}}{1-e^{i\frac{4kqp\pi}{N}}}(1-e^{i\frac{4kqp\pi
N}{N}})
\end{eqnarray*} and $\gcd(pq,N)=1$ and $\gcd(pr,2)=1$.

When $0 < k < N$, if $1-e^{i\frac{4kpq\pi}{N}}=0$, then $2k=N$, the
real part of

$$e^{i\frac{krp\pi}{2N}}+e^{i\frac{k(4q+r)p\pi}{N}}+\cdots+e^{i\frac{k[4q(N-1)+r]p\pi}{N}}$$
is zero.

If $1-e^{i\frac{4kpq\pi}{N}}\neq0$, then
$$e^{i\frac{krp\pi}{N}}+e^{i\frac{k(4q+r)p\pi}{N}}+\cdots+e^{i\frac{k[4q(N-1)+r]p\pi}{N}}=0$$

From equality (\ref{formula020}), when  $0< k_1<N$, the inner
product of the $k_1$th row and itself of $C(N)$ is 1.  We know that
the product of $C(N)$ and its transpose is an identity matrix. Thus,
it is easy to get the inverse transform of transform
(\ref{formula01}). We omit it here.\\

We remark the following three points:

(I) The transpose of $C(N)$ is an orthogonal matrix, so the
transform below is also orthogonal.
\begin{eqnarray*}
X(k)=\sum\limits^{N-1}_{n=0}\alpha(n)x(n)\cos\frac{n(4qk+r)p\pi}{2N},\\
k=0,1,\cdots,N-1.
\end{eqnarray*}
where $\gcd(pq,N)=1$, $\gcd(pr,2)=1$, $p$, $q$ and $r$ are positive
integers.

(II)    We can obtain a generalized DCT-IV-E transform, as follows:
\begin{eqnarray*}
X(k)=\alpha(k)\sum\limits^{N-1}_{n=0}x(n)\cos\frac{(2k+1)(4qn+r)p\pi}{4N},\\
k=0,1,\cdots,N-1.
\end{eqnarray*}
where $\gcd(pq,N)=1$, $\gcd(pr,2)=1$, $p$, $q$ and $r$ are positive
integers.

(III)  In a similar way, we can generalize discrete sine transforms,
such as DST-II-E, DST-III-E, DST-IV-E.\\

\section{A new type of discrete cosine transform}

Let $\{x(n);n=0,1,2,\cdots$, $N-1\}$ be a vector of real numbers. We
define a new form of discrete cosine transform,  as
follows:

{\scriptsize
\begin{eqnarray}
X(k)=\sqrt{\frac{4}{2N-1}}\alpha(k)\sum\limits^{N-1}_{n=0}\alpha(n)x(n)\cos\frac{(2k+1)(2n+1)\pi}{2N-1},\label{formula03}\\
 k=0,1,\cdots,N-1,\notag
\end{eqnarray}}
where
\begin{eqnarray*}
&&\alpha(k)=\left\{
\begin{array}{ll}
\frac{1}{\sqrt 2} &             k=N-1\\
\ \ 1               &            \mbox{else}\\
\end{array}\right.\\
&&\alpha(n)=\left\{
\begin{array}{ll}
\frac{1}{\sqrt 2} &             n=N-1\\
\ \ 1               &          \mbox{else}\\
\end{array}\right.
\end{eqnarray*}
The transform above can be written into matrix form:
\begin{eqnarray*}
X(N)=C(N)\cdot x(N)
\end{eqnarray*}
where $X(N),\ x(N)$ are column vectors of length $N,$

{\tiny $C(N)=\sqrt{\frac{4}{2N-1}}\\
\left(
\begin{array}{ccccc}
\cos\frac{\pi}{2N-1}   &\cos\frac{3\pi}{2N-1}  &\cdots &\cos\frac{(2N-3)\pi}{2N-1}    &-\frac{1}{\sqrt2}\\
\cos\frac{3\pi}{2N-1}   &\cos\frac{9\pi}{2N-1}  &\cdots &\cos\frac{3(2N-3)\pi}{2N}    &-\frac{1}{\sqrt2}\\
\vdots&    \vdots      &\vdots&   \vdots&  \vdots \\
\cos\frac{(2N-3)\pi}{2N-1}   &\cos\frac{(2N-3)3\pi}{2N-1}  &\cdots &\cos\frac{(2N-3)(2N-3)\pi}{2N-1}   &-\frac{1}{\sqrt2}\\
-\frac{1}{\sqrt2}    &-\frac{1}{\sqrt2}    &\cdots &-\frac{1}{\sqrt2} & -\frac{1}{2}\\
\end{array}\right)$
}\\\\

 For the sake of simplicity, the coefficient
 $\sqrt{\frac{4}{2N-1}}$   is omitted in the following discussions. Let $0\leq k_1<k_2<N-1$, the inner product of the $k_1$th row and
$k_2$th row of $C(N)$ is that£º {\scriptsize
\begin{eqnarray}
&&\cos\frac{(2k_1+1)\pi}{2N-1}\cos\frac{(2k_2+1)\pi}{2N-1}\notag\\
&&+\cos\frac{(2k_1+1)3\pi}{2N-1}\cos\frac{(2k_2+1)3\pi}{2N-1}+\cdots\notag\\
&&+\cos\frac{(2k_1+1)(2N-3)\pi}{2N-1}\cos\frac{(2k_2+1)(2N-3)\pi}{2N-1}+\frac{1}{2}\notag\\
&=&\frac{1}{2}[\cos\frac{2(k_1-k_2)\pi}{2N-1}+\cos\frac{2(k_1+k_2+1)\pi}{2N-1}\notag\\
&&+\cos\frac{2(k_1-k_2)3\pi}{2N-1}+\cos\frac{2(k_1+k_2+1)3\pi}{2N-1}+\cdots\notag\\
&&+\cos\frac{2(k_1-k_2)(2N-3)\pi}{2N-1}\notag\\
&&+\cos\frac{2(k_1+k_2+1)(2N-3)\pi}{2N-1}]+\frac{1}{2}\label{formula04}
\end{eqnarray}}

Since  $e^{ix}=\cos x+i\sin x$, we consider the real part of the
following complex number:
\begin{eqnarray*}
&&e^{i\frac{2k\pi}{2N-1}}+e^{i\frac{2k3\pi}{2N-1}}+\cdots+e^{i\frac{2k(2N-3)\pi}{2N-1}}\\
&=&
\frac{e^{i\frac{2k\pi}{2N-1}}}{1-e^{i\frac{4k\pi}{2N-1}}}(1-e^{i\frac{4k(N-1)\pi}{2N-1}})
\end{eqnarray*}

When $0<k<2N-1,1-e^{i\frac{4k\pi}{2N-1}}\neq0$.

Note that
\begin{eqnarray*}
&& e^{i\frac{4k\pi(N-1)}{2N-1}}\\
&=&\cos(2k\pi-\frac{2k\pi}{2N-1})+i\sin(2k\pi-\frac{2k\pi}{2N+1})\\
&=&\cos(\frac{2k\pi}{2N-1})-i\sin(\frac{2k\pi}{2N-1})
\end{eqnarray*}

By setting $\alpha=\frac{2k\pi}{2N-1}$, we have
\begin{eqnarray*}
&&\frac{e^{i\frac{2k\pi}{2N-1}}}{1-e^{i\frac{4k\pi}{2N-1}}}\\
&=&\frac{\cos\alpha+i\sin\alpha}{1-\cos2\alpha-i\sin2\alpha}\\
&=&\frac{\cos\alpha+i\sin\alpha}{2\sin\alpha(\sin\alpha-i\cos\alpha)}\\
&=&\frac{(\cos\alpha+i\sin\alpha)(\sin\alpha+i\cos\alpha)}{2\sin\alpha(\sin\alpha-\cos\alpha)(\sin\alpha+i\cos\alpha)}\\
&=&\frac{i}{2\sin\alpha}
\end{eqnarray*}

Thus, the real part of
 $$e^{i\frac{2k\pi}{2N-1}}+e^{i\frac{2k3\pi}{2N-1}}+\cdots+e^{i\frac{2k(2N-3)\pi}{2N-1}}$$
is $\frac{i}{2\sin(\alpha)}i\sin(\alpha)=-\frac{1}{2}$.\\

From equality (\ref{formula04}), for $0\leq k_1<k_2<N-1$, the inner
product of the $k_1$th row and $k_2$th row of $C(N)$ is

$$\frac{1}{2}(-\frac{1}{ 2}-\frac{1}{2})+\frac{1}{2}=0$$

 By setting
$\alpha=\frac{(2k+1)\pi}{2N-1}$, we have
\begin{eqnarray*}
&&e^{i\frac{(2k+1)\pi}{2N-1}}+e^{i\frac{(2k+1)3\pi}{2N-1}}+\cdots+e^{i\frac{(2k+1)(2N-3)\pi}{2N-1}}\\
&=&\frac{e^{i\frac{(2k+1)\pi}{2N-1}}}{1-e^{i\frac{2(2k+1)\pi}{2N-1}}}(1-e^{i\frac{2(2k+1)(N-1)\pi}{2N-1}})\\
&=&\frac{i}{2\sin\alpha}(1+\cos\alpha-i\sin\alpha)\\
&=&\frac{1}{2}+\frac{i}{2\sin\alpha}(1+\cos\alpha)
\end{eqnarray*}

Now, for $0\leq k_1<N-1$, the inner product of the $k_1$th row and
$(N-1)$st row of $C(N)$ is

$$\frac{1}{2}(-\frac{1}{\sqrt 2})+(-\frac{1}{2})(-\frac{1}{\sqrt 2})=0$$

Thus,  $C(N)$ is an orthogonal matrix. Equivalently,  transform (\ref{formula03}) is orthogonal.

It is easy to know that the product of $C(N)$ and its transpose is
an identity matrix. Thus, we can get the inverse transform of
transform
(\ref{formula03}). We omit it here.\\

Similarly, we can obtain a new form of discrete sine transform as
follows:
\begin{eqnarray*}
X(k)=\sqrt\frac{4}{2N+1}\sum\limits^{N-1}_{n=0}x(n)\sin{\frac{(2k+1)(2n+1)\pi}{2N+1}},\\
k=0,1,\cdots,N-1.
\end{eqnarray*}

And a  new form of discrete sine-cosine transform as follows:
\begin{eqnarray*}
X(k)=\sqrt{{\frac{2}{2N+1}}}\sum\limits^{2N}_{n=0}x(n)\cos{\frac{(2k+1)(2n+1)\pi}{2N+1}},\\
k=0,1,\cdots,N-1;
\end{eqnarray*}
\begin{eqnarray*}
X(N)=-\sqrt{{\frac{1}{2N+1}}}\sum\limits^{2N}_{n=0}x(n);
\end{eqnarray*}
\begin{eqnarray*}
X(N+1+k)=\sqrt{\frac{2}{2N+1}}\sum\limits^{2N}_{n=0}x(n)\sin\frac{(2k+1)(2n+1)\pi}{2N+1},\\
k=0,1,\cdots,N-1 .
\end{eqnarray*}

The orthogonality of above transforms  follows from an analysis
similar to that of new discrete cosine transform
(\ref{formula03}).\\

\section{Generalized discrete W transform}

 Bracewell  (1983)%\cite{Bracewell}
advanced a discrete Hartlry transform (DHT) in the domain of real
numbers. Its nuclear function $\mbox{cas }wt=\cos wt+\sin wt$ is the
sum of the real part and imaginary part of  Fourier transform
nuclear function $\exp(iwt)=\cos wt+i \sin wt$. Zhongde Wang (1984)%\cite{Wang84} 
generalized DHT and advanced discrete W transform
(DWT), which is an unified form with two parameters. There are four
meaningful types (including DHT) in DWT. The nuclear function of DWT
is still $\mbox{cas } wt=\cos wt+\sin wt$. The discrete W transform
has been used in frequency spectrum analysis, data compression,
convolution computation and information hiding. Its theory and
algorithms have received much attention for the last two decades
(Wang,1989, Wang,1992,  Kunz,2008, Zhang,1992).

Let $\{x(n);n=0,1,2,\cdots$, $N-1\}$ be a vector of real numbers.
The unified form of discrete W transform is defined
(Wang,1984,Wang,1985) as follows:

\begin{eqnarray*}
X(k)=\sqrt{\frac{2}{N}}\sum\limits^{N-1}_{n=0}x(n)\sin[\frac{\pi}{4}+(k+\alpha)(n+\beta)\frac{2\pi}{N}],\\
k=0,1,\cdots,N-1
\end{eqnarray*}

The transform above  is orthogonal only for $(\alpha,\beta)\in
\{(0,0)$,$(\frac{1}{2},0)$,$(0,\frac{1}{2})$,$(\frac{1}{2},\frac{1}{2})\}$.
 It is discrete Hartley transform when $(\alpha,\beta)=(0,0)$. The four
transforms are defined as DWT-I, DWT-II, DWT-III and DWT-IV,
respectively.

Actually, the unified form of discrete W transform can be added with
one more  parameter $\gamma$,  as follows:
\begin{eqnarray}
X(k)=\sqrt{\frac{2}{N}}\sum\limits^{N-1}_{n=0}x(n)\sin[\frac{\pi}{4}+(k+\alpha)(n+\beta)\frac{\gamma\pi}{N}],
\label{formula01}\\
k=0,1,\cdots,N-1 \notag
\end{eqnarray}

This is the proof that when
$(\alpha,\beta,\gamma)=(\frac{1}{2},\frac{r}{q},2pq)$ and
$\gcd(pq,N)=1$, where
 $p$,  $q$ and $r$ are  positive integers,  transform (\ref{formula01}) is
 orthogonal.

For the sake of simplicity, when
$(\alpha,\beta,\gamma)=(\frac{1}{2},\frac{r}{q},2pq)$ and
$\gcd(pq,N)=1$,  transform (\ref{formula01}) is rewritten as
follows:

\begin{eqnarray}
X(k)=\frac{1}{\sqrt{N}}\sum\limits^{N-1}_{n=0}x(n)\mbox{cas}[(2k+1)(qn+r)\frac{p\pi}{N}],\label{formula02}\\
k=0,1,\cdots,N-1, \notag
\end{eqnarray}
where $\mbox{cas }x=\cos x+\sin x$.\\

 Transform (\ref{formula02}) can be written into matrix form:
\begin{eqnarray*}
X(N)=C(N)x(N)
\end{eqnarray*}
where $X(N),x(N)$ are column vectors of length $N,$

%\onecolumn

{\tiny $ H(N)=\frac{1}{\sqrt{N}}\\ \left(
\begin{array}{cccccllllll}
\mbox{cas}\frac{rp\pi}{N}    &\mbox{cas}\frac{(q+r)p\pi}{N}    &\mbox{cas}\frac{(2q+r)p\pi}{N}  &\cdots  &\mbox{cas}\frac{[q(N-1)+r]p\pi}{N}\\
\mbox{cas}\frac{3rp\pi}{N}   &\mbox{cas}\frac{3(q+r)p\pi}{N}   &\mbox{cas}\frac{3(2q+r)p\pi}{N} &\cdots  & \mbox{cas}\frac{3[q(N-1)+r]p\pi}{N}\\
\mbox{cas}\frac{5rp\pi}{N}  &\mbox{cas}\frac{5(q+r)p\pi}{N}   &\mbox{cas}\frac{5(2q+r)p\pi}{N} &\cdots  &\mbox{cas}\frac{5[q(N-1)+r]p\pi}{N}\\
\vdots                &\vdots       &\vdots &\vdots &\vdots\\
\mbox{cas}\frac{r\eta}{N} &\mbox{cas}\frac{(q+r)\eta}{N} &\mbox{cas}\frac{(2q+r)\eta}{N} &\cdots &\mbox{cas}\frac{[q(N-1)+r]\eta}{N}\\
\end{array}\right)$
}\\\\
where $\eta=(2N-1)p\pi$.

To prove that   transform (\ref{formula02}) is orthogonal,  or
equivalently, to prove that $H(N)$ is an orthogonal matrix.

It is easy to know that

\begin{eqnarray*}
&&\mbox{cas}A \mbox{cas}B\\&=&(\cos A+\sin A)(\cos B+\sin B)\\
&=& \cos A\cos B+\sin A\sin B+\sin A\cos B+\cos A\sin B\\
&=& \cos (A-B)+\sin (A+B)
\end{eqnarray*}

To simplify  the problem, we omit the coefficient $\frac{1}{\sqrt
N}$. For $1\le k_1<k_2\leq N$, the inner product of the $k_1$th row
and $k_2$th row of $H(N)$ is that£º

{\scriptsize
\begin{eqnarray}
&&\mbox{cas}\frac{(2k_1-1)rp\pi}{N}\mbox{cas}\frac{(2k_2-1)rp\pi}{N}\notag\\
&&+\mbox{cas}\frac{(2k_1-1)(q+r)p\pi}{N}\mbox{cas}\frac{(2k_2-1)(q+r)p\pi}{N}
+\cdots\notag\\
&&+\mbox{cas}\frac{(2k_1-1)[q(N-1)+r]p\pi}{N}\mbox{cas}\frac{(2k_2-1)[q(N-1)+r]p\pi}{N}\notag\\
&=& \cos \frac{2(k_1-k_2)rp\pi}{N}+\sin
\frac{2(k_1+k_2-1)rp\pi}{N}\notag\\
&&+\cos \frac{2(k_1-k_2)(q+r)p\pi}{N}+\sin
\frac{2(k_1+k_2-1)(q+r)p\pi}{N}\notag\\
&&+\cos \frac{2(k_1-k_2)[q(N-1)+r]p\pi}{N}\notag\\
&&+\sin \frac{2(k_1+k_2-1)[q(N-1)+r]p\pi}{N}\label{formula03}
\end{eqnarray}
}

As for the function $e^{ix}=\cos x+i\sin x$, we consider
$$e^{i\frac{2krp\pi}{N}}+e^{i\frac{2k(q+r)p\pi}{N}}+\cdots
+e^{i\frac{2k[q(N-1)+r]p\pi}{N}}$$ where $0< k<2N$. Note that
\begin{eqnarray*}
&&e^{i\frac{2krp\pi}{N}}+e^{i\frac{2k(q+r)p\pi}{N}}+\cdots+e^{i\frac{2k[q(N-1)+r]p\pi}{N}}\\
&=&\frac{e^{i\frac{2krp\pi}{N}}}{1-e^{i\frac{2kqp\pi}{N}}}(1-e^{i\frac{2kqp\pi
N}{N}})
\end{eqnarray*}

When $0<k<N$ and $\gcd(pq,N)=1$, we know that
$1-e^{i\frac{2kqp\pi}{N}}\neq0$. Since $$1-e^{i\frac{2kqp\pi
N}{N}}=1-e^{i2kqp\pi}=0,$$ hence
$$e^{i\frac{2krp\pi}{N}}+e^{i\frac{2k(q+r)p\pi}{N}}+\cdots+e^{i\frac{2k[q(N-1)+r]p\pi}{N}}=0$$

Thus
\begin{eqnarray*}
\cos\frac{2(k_1-k_2)rp\pi}{N}+\cos\frac{2(k_1-k_2)(q+r)p\pi}{N}+\\
\cdots+\cos\frac{2(k_1-k_2)[q(N-1)+r]p\pi}{N}=0
\end{eqnarray*}

When $0<k<2N$, if $1-e^{i\frac{2kqp\pi}{N}}=0$, then $k=N$, thus,
the imaginary part of
$$e^{i\frac{2krp\pi}{N}}+e^{i\frac{2k(q+r)p\pi}{N}}+\cdots+e^{i\frac{2k[q(N-1)+r]p\pi}{N}}$$ is
zero.

When $0<k<2N$ and if $1-e^{i\frac{2kqp\pi}{N}}\neq0$, then
$$e^{i\frac{2krp\pi}{N}}+e^{i\frac{2k(q+r)p\pi}{N}}+\cdots+e^{i\frac{2k[q(N-1)+r]p\pi}{N}}=0$$

Thus, {\scriptsize
\begin{eqnarray*}
\sin \frac{2(k_1+k_2-1)rp\pi}{N}+\sin \frac{2(k_1+k_2-1)(q+r)p\pi}{N}+\cdots\\
+\sin \frac{2(k_1+k_2-1)[q(N-1)+r]p\pi}{N}=0
\end{eqnarray*}
}

From equality (\ref{formula03}),  the inner product of the $k_1$th
row and $k_2$th row of $H(N)$ is zero for $1\le k_1<k_2\leq N$,
namely $H(N)$ is an orthogonal
matrix.\\

For $1\le k_1\leq N$, the inner product of the $k_1$th row  of
$H(N)$ and itself is that

 {\small \begin{eqnarray*}
&&\frac{1}{N}\{[\mbox{cas}\frac{(2k_1-1)rp\pi}{N}]^2+[\mbox{cas}\frac{(2k_1-1)(q+r)p\pi}{N}]^2+\cdots\\
&&+[\mbox{cas}\frac{(2k_1-1)[q(N-1)+r]p\pi}{N}]^2\}\\
&=& 1+\frac{1}{N}\{\sin \frac{2(2k_1-1)rp\pi}{N}+\sin
\frac{2(2k_1-1)(q+r)p\pi}{N}\\
&&+\cdots+\sin \frac{2(2k_1-1)[q(N-1)+r]p\pi}{N}\}
\end{eqnarray*}
}

Note that $0<2k_1-1<2N$, from the proof above, we know that
\begin{eqnarray*}\sin \frac{2(2k_1-1)rp\pi}{N}+\sin
\frac{2(2k_1-1)(q+r)p\pi}{N}\\+\cdots+ \sin
\frac{2(2k_1-1)[q(N-1)+r]p\pi}{N}=0\end{eqnarray*}

It follows that  the inner product of the $k_1$th row  of $H(N)$ and
itself is 1 for $1\le k_1\leq N$, namely the
 product of $H(N)$ and its transpose is an identity
matrix.   Thus, it is easy to get the inverse transform of transform
(\ref{formula02}),

\begin{eqnarray*}
x(k)=\frac{1}{\sqrt{N}}\sum\limits^{N-1}_{n=0}X(n)\mbox{cas}[(2n+1)(qk+r)\frac{p\pi}{N}],\\
k=0,1,\cdots,N-1,
\end{eqnarray*}
where $\mbox{cas }x=\cos x+\sin x$.\\

We should remark the following two points:

 (I)  There is no limit to
parameter $r$ in the new transform  (2), $r$ is an arbitrary
positive integer.

 (II)  In a similar way, we can generalize DWT-IV  as follows:
\begin{eqnarray*}
X(k)=\frac{1}{\sqrt{N}}\sum\limits^{N-1}_{n=0}x(n)\mathrm{cas}[(2k+1)(2qn+r)\frac{p\pi}{2N}],\\
k=0,1,\cdots,N-1
\end{eqnarray*}
where $\gcd(pq,N)=1$, $p$, $q$ and $r$ are positive integers. It is
 degraded to transform (2) when $r$ is an even number.\\

Consider two special cases of transform (\ref{formula01}):

  When
$(\alpha,\beta,\gamma)=(\frac{1}{2},\frac{1}{2},4)$ and $N$ is an
odd number, transform (\ref{formula01}) can be written as follows:
\begin{eqnarray*}
X(k)=\frac{1}{\sqrt{N}}\sum\limits^{N-1}_{n=0}x(n)\mbox{cas}\frac{(2k+1)(2n+1)\pi}{N},\\
k=0,1,\cdots,N-1 \end{eqnarray*}

When $(\alpha,\beta,\gamma)=(\frac{1}{2},\frac{1}{q},2q)$ and $\gcd(
q, N) =1$, transform (\ref{formula01}) can be written as follows:
\begin{eqnarray*}
X(k)=\frac{1}{\sqrt{N}}\sum\limits^{N-1}_{n=0}x(n)\mbox{cas}\frac{(2k+1)(q n+1)\pi}{2N},\\
k=0,1,\cdots,N-1
\end{eqnarray*}

We know that transforms above are orthogonal.
\\

\section*{ Acknowledgment}
The research was supported by Zhejiang Natural Science Foundation (No.Y1100318, R1090138) and Chinese Natural Science Foundation
(No. 60802047).

{\bf Jianqin Zhou} received his B.Sc. degree in mathematics from
East China Normal University, China, in 1983, and M.Sc. degree in
probability and statistics from Fudan University, China, in 1989.
From 1989 to 1999 he was with the Department of Mathematics and
Computer Science, Qufu Normal University, China. From 2000 to 2002,
he worked for a number of IT companies in Japan. From 2003 to 2007
he was with the Department of Computer Science, Anhui University of
Technology, China.
 From Sep 2006  to
Feb 2007, he was a visiting scholar with the Department of
Information and Computer Science, Keio University, Japan. Since 2008
he has been with the Telecommunication School, Hangzhou Dianzi
University, China

He  published more than 70 papers, and  proved a conjecture posed by
famous mathematician Paul Erd\H{o}s et al. His research interests
include coding theory, cryptography and combinatorics.

\end{document}